\documentclass[conference]{IEEEtran}
\usepackage[T1]{fontenc}
\usepackage{psfrag,amsmath,amssymb,color,cite, amsmath,graphicx}
\usepackage{amsthm}
\usepackage{listings}
\usepackage{stfloats}
\usepackage{mathtools}

\newcommand{\DeclareAutoPairedDelimiter}[3]{%
  \expandafter\DeclarePairedDelimiter\csname Auto\string#1\endcsname{#2}{#3}%
  \begingroup\edef\x{\endgroup
    \noexpand\DeclareRobustCommand{\noexpand#1}{%
      \expandafter\noexpand\csname Auto\string#1\endcsname*}}%
  \x}
\DeclareAutoPairedDelimiter\modulo{[}{]} 
\usepackage{fancyhdr,graphicx,amsmath,amssymb}
\usepackage[linesnumbered,ruled,vlined]{algorithm2e}
\include{pythonlisting}
\usepackage{mathrsfs}
\usepackage{multirow}
\usepackage{pifont}
\usepackage[colorinlistoftodos]{todonotes}

\newcommand{\beqn}{\begin{equation}}
\newcommand{\eeqn}{\end{equation}}
\newcommand{\beqa}{\begin{eqnarray}}
\newcommand{\eeqa}{\end{eqnarray}}
\newcommand{\beqas}{\begin{eqnarray*}}
\newcommand{\eeqas}{\end{eqnarray*}}

\presetkeys{todonotes}{color=green!30}{}
\usepackage{amsmath}
\usepackage{amsfonts}
\usepackage{graphicx}
\usepackage{array}
\usepackage{ amssymb }
\newcolumntype{P}[1]{>{\centering\arraybackslash}p{#1}}
\newcolumntype{M}[1]{>{\centering\arraybackslash}m{#1}}

\newcommand{\thickbar}[1]{\mathbf{\bar{\text{$#1$}}}}
\newcommand{\thicktilde}[1]{\mathbf{\widetilde{\text{$#1$}}}}

\makeatletter
\def\blfootnote{\xdef\@thefnmark{}\@footnotetext}
\makeatother
\begin{document}

\title{ Unitary-Precoded Single-Carrier Waveforms
for High Mobility: Detection and Channel Estimation
}
\author{\IEEEauthorblockN{Tharaj Thaj and Emanuele Viterbo}\\
\IEEEauthorblockA{ECSE Department, Monash University, Clayton, VIC 3800, Australia\\
Email: \{tharaj.thaj, emanuele.viterbo\}@monash.edu}}

\maketitle
\begin{abstract}
This paper presents unitary-precoded single-carrier (USC) modulation as a family of waveforms based on multiplexing the information symbols on time domain unitary basis functions. The common property of these basis functions is that they span the entire time and frequency plane.
The recently proposed orthogonal time frequency space (OTFS) and orthogonal time sequency multiplexing (OTSM)  based on discrete Fourier transform (DFT) and Walsh Hadamard transform (WHT), respectively, fall in the general framework of USC waveforms. In this work, we present channel estimation and detection methods that work for any USC waveform and numerically show that any choice of unitary precoding results in the same error performance. Lastly, we implement some USC systems and compare their performance with OFDM in a real-time indoor setting using an SDR platform. 
\end{abstract}

\section{Introduction}
\blfootnote{This work was supported by the Australian Research Council through the Discovery Project
under Grant DP200100096.}
Reliable communications in high mobility channels are a key promise of 6G wireless communications.  With the advent of high-speed trains, unmanned aerial vehicles (UAVs), and drones, there is an urgent need to address the issue of reliable communication in high-mobility wireless channels. Widely used modulation schemes such as orthogonal frequency division multiplexing (OFDM) multiplex information symbols in the time-frequency domain. The advantage of such data transmission is that the orthogonality property enables the use of a single tap equalizer to detect the transmitted data at the receiver, thereby providing a low complexity solution to reliable communication in frequency selective (static) channels. However, OFDM suffers from performance degradation in time-frequency selective channels due to the loss of orthogonality and unequal SNR of its sub-carriers.   

Recently, orthogonal time frequency space (OTFS) modulation was proposed, which multiplexes information symbols in the delay-Doppler domain, \cite{Hadani,Elsevier_book}. The key advantage compared to OFDM is that all the information symbols experience a roughly flat fading channel. OTFS achieves this by spreading all the information symbols on unitary basis functions that span the entire available time and bandwidth resource using the inverse symplectic fast Fourier transform (ISFFT) operation, thereby exploiting maximum time-frequency diversity. It was proved in \cite{GOP} that any {\em constant modulus} 2-D unitary transformation in the time-frequency domain could offer the same performance as OTFS.

However, 2-D precoding in the time-frequency domain may increase the transceiver complexity since an OFDM modulator on top of 2-D precoding is still required for transmission, \cite{OTSM_conf}. One solution to mitigating the 2-D  time-frequency precoding complexity in \cite{GOP} is to restrict the unitary transformation precoding along the frequency dimension to the Fourier transform, specifically FFT as in the case of OTFS. Such precoding along the frequency dimension transforms the time-frequency information samples directly into the delay-time domain, thereby bypassing the need for an OFDM modulator. Using FFT as above, the 2-D precoding along the frequency-time domain is reduced to just a 1-D precoding along the time dimension of the delay-time domain.

Since any arbitrary unitary matrix can be used for precoding, every choice leads to a new waveform with the same error performance. Therefore, it seems reasonable to group such waveforms under a general framework, which we refer to as unitary-precoded single-carrier (USC) waveforms in this work. With such motivation, in this work, we first generalize OTFS and OTSM (proposed in \cite{OTSM_journ,OTSM_conf}) as a USC waveform with DFT and WHT precoding, respectively, along the time dimension of the delay-time domain. We then present time-domain channel estimation and detection methods that work for any USC waveform and numerically show that any choice of unitary precoding results in the same error performance. Even though OTFS has gained popularity recently, other USC waveforms like OTSM exist that can outperform OTFS for key performance requirements such as transceiver complexity and ease of implementation. Lastly, USC with DFT and WHT unitary matrices are implemented and tested in real-time on an SDR platform to validate its' superior performance over OFDM even in low-mobility channels.

{\em Notations}:
The following notations will be followed in this paper: $a$, $\bf{a}$, ${\bf A}$ represent a scalar, vector, and matrix, respectively; ${\bf a}[n]$ and ${\bf A}[m,n]$ represent the $n$-th and $(m,n)$-th element of ${\bf a}$ and ${\bf A}$, respectively; ${\bf A}^{\rm T}$, ${\bf A}^\dag$, ${\bf A}^*$ and ${\bf A}^n$ represent the transpose, Hermitian transpose, complex conjugate and $n$-th power, respectively, of ${\bf A}$. 
The set of $M \times N$ dimensional matrices with complex entries are denoted by ${\mathbb{C}}^{M \times N}$. 
Let $\otimes$ denote the Kronecker product, $|\mathcal{S}|$ denote the cardinality of the set $\mathcal{S}$, vec$({\bf A})$, the column-wise vectorization of the matrix ${\bf A}$ and ${\rm vec}_{N,M}^{-1}({\bf a})$ is the matrix formed by folding a vector ${\bf a}$ into a $N\times M$ matrix by filling it column wise. Let ${\bf F}_N$ be the normalized $N$ point DFT matrix with elements ${\bf F}_N(i,k)=(1/\sqrt{N}){\rm e}^{-j2\pi ik/N}$ and ${\bf W}_N$ be the normalized $N$-point WHT matrix, \cite{Hadamard_text},
\section{System Model}
We will be using the following matrix/vector representation throughout the paper. Let ${\bf x}, {\bf y} \in \mathbb{C}^{NM \times 1}$ be the transmitted and received information symbols. The total frame duration and bandwidth of the transmitted signal frame are $T_f=NT$ and $B = M \Delta f$, respectively. We consider the case where $T\Delta f=1$, i.e., the signal is critically sampled for any pulse shaping waveform.
\subsection{ Conventional multi-carrier transmission}
Let ${\bf X} \in \mathbb{C}^{M \times N}$ be the 2-D information symbol matrix. Let ${\bf X}_{\rm FT}\in \mathbb{C}^{M \times N}$ be the frequency-time samples generated by precoding the information symbols in ${\bf X}$. ${\bf X}_{\rm FT}(m,n)$ represents the precoded data sample transmitted in the $m$-th sub-carrier of the $n$-th time slot, where $m=0,\ldots,M-1$ and $n=0,\ldots,N-1$. Let $\thicktilde{\bf X} \in \mathbb{C}^{M\times N}$ be the delay-time samples generated from the time-frequency samples using the Heisenberg transform as
\begin{equation}
    \thicktilde{\bf X}={\bf G}_{\rm TX}\cdot{\bf F}_M^{\dag}\cdot{\bf X}_{\rm FT} \label{FT_to_DT}
\end{equation}
where ${\bf G}_{\rm TX}$ is the pulse shaping waveform at the transmitter. We assume the rectangular pulse shaping waveform in this paper, i.e., ${\bf G}_{\rm TX}={\bf I}_M$. The operation is (\ref{FT_to_DT}) is the conventional OFDM modulator. The $NM \times 1$ time domain samples are generated by vectorizing the delay-time domain samples as
\begin{equation}
{\bf s}={\rm vec}(\thicktilde{\bf X}) \label{DT_to_time}
\end{equation}
The time-domain signal are divided into $N$ blocks ${\bf s}_n$ of size $M$ as ${\bf s}=[{\bf s}_0^{\rm T},\ldots,{\bf s}_{N-1}^{\rm T}]^{\rm T}$ and finally a guard interval of length $L_{\rm G}$ is added to each time domain block in ${\bf s}$ to avoid inter-block interference. The guard intervals can either be filled with a cyclic prefix (CP) or zero-padding (ZP). 
\subsection{Generalized 2-D unitary frequency-time precoding}
The generalized 2-D unitary precoding  can be written in matrix form as
\begin{equation}
    {\bf X}_{\rm FT}={\bf U}_{\rm F}\cdot{\bf X}\cdot{\bf U}_{\rm T}
\end{equation}
where ${\bf U}_{\rm F} \in \mathbb{C}^{M \times M}$ and ${\bf U}_{\rm T} \in \mathbb{C}^{N \times N}$ are the precoding matrices along the frequency and time domain, respectively.

In \cite{GOP}, it was proved that any orthogonal 2-D transformation with {\em constant modulus} basis functions operating on the entire frequency-time domain allows the receiver to exploit maximum frequency-time diversity in doubly-selective channels. 
This implies that, in terms of error performance, the best choice for ${\bf U}_{\rm F}$ and ${\bf U}_{\rm T}$ are unitary transforms, such as DFT and WHT. The precoding matrices ${\bf U}_{\rm F}$ and ${\bf U}_{\rm T}$ for different modulation schemes are listed in Table \ref{table1}.
\begin{table}
\begin{center}
\caption{${\bf U}_{\rm F}$ and ${\bf U}_{\rm T}$ representing different modulation schemes}
\begin{tabular}{ | m{9em} | m{1cm}| m{1cm} | } 
\hline
Modulation scheme & ${\bf U}_{\rm F}$ & ${\bf U}_{\rm T}$ \\ 
\hline
OFDM  & ${\bf I}_M$ & ${\bf I}_N$ \\
\hline
SC  & ${\bf F}_M$ & ${\bf I}_N$ \\
\hline
OTFS & ${\bf F}_M$ & ${\bf F}_N^{\dag}$ \\ 
\hline
OTSM & ${\bf F}_M$ & ${\bf W}_N$ \\ 
\hline
\end{tabular}
\label{table1}
\end{center}
\vspace{-6mm}
\end{table}
\subsection{Generalized 1-D unitary time precoding}
A 2-D unitary precoding in the frequency-time domain may increase the complexity of transceiver modulation and demodulation. One way to mitigate this complexity is by restricting the unitary transformation along the frequency domain to the FFT, i.e., ${\bf U}_{\rm F}={\bf F}_M$,
\begin{equation}
    {\bf X}_{\rm FT}={\bf F}_M\cdot{\bf X}\cdot{\bf U}_{\rm T}
\end{equation}
This simplifies the Heisenberg transform in (\ref{FT_to_DT}) as (assuming practical rectangular pulse shaping waveforms):
\begin{equation}
    \thicktilde{\bf X}={\bf F}_M^{\dag}\cdot({\bf F}_M\cdot{\bf X}\cdot{\bf U}_{\rm T})={\bf X}\cdot{\bf U}_{\rm T} \label{FT_to_DT2}
\end{equation}
Now, (\ref{FT_to_DT2}) can be considered as a precoding along the time dimension of the delay-time domain, i.e., the rows of $\thicktilde{\bf X}$. 
The time-domain signal is then generated as 
\begin{equation}
    {\bf s}={\rm vec}(\thicktilde{\bf X}) \label{vec}
\end{equation}
The operations in (\ref{FT_to_DT2}) and (\ref{vec}) can be combined as
\begin{equation}
    {\bf s}[m+kM]=\thicktilde{\bf X}[m,k]=\sum_{n=0}^{N-1}{\bf X}[m,n]{\bf U}_{\rm T}[n,k] \label{VT_Zak}
\end{equation}
 for $m=0,\ldots M-1$ and $k=0,\ldots,N-1$. This 1-D unitary precoding can still achieve maximum time-frequency diversity similar to a 2-D unitary precoding in the frequency-time domain, but with much lower modulation/demodulation complexity. For ${\bf U}_{\rm T}={\bf F}_N^{\dag}$, the operation in (\ref{VT_Zak}) is known in the literature as the inverse discrete Zak transform (IDZT) and is equivalent to the OTFS transmitter for rectangular pulse shaping waveforms, \cite{Journ_OTFS,Saif}. 
\subsection{USC Transmitter}
Let ${\bf X} \in \mathbb{C}^{M \times N}$ be the 2-D information symbols. The transmitted information symbol matrix ${\bf X}$ and the delay-time samples are vectorized as
\begin{equation}
    {\bf x}={\rm vec}({\bf X}^{\rm T}),\quad\thicktilde{\bf x}={\rm vec}(\thicktilde{\bf X}^{\rm T})
\end{equation}
Let ${\bf P}$ be the row-column interleaver matrix which writes the $NM$ samples column-wise into a $N \times M$ matrix and reads the elements out row-wise. The time-domain samples can then be generated using (\ref{VT_Zak}) as
\begin{equation}
    {\bf s}={\bf P}\cdot({\bf I}_M\otimes {\bf U}_{\rm T})\cdot{\bf x}={\bf P}\cdot\thicktilde{\bf x} \label{tx_perm}
\end{equation}
where ${\bf U}_{\rm T}$ is the precoding matrix along the time-domain given for different waveforms in the literature as given in Table \ref{table1}.
 The time domain samples are then pulse shaped, digital to analog converted and transmitted into the wireless medium as $s(t)$ at a carrier frequency $f_c$. 
\subsection{Channel}
Consider a channel with $P$ paths, where $h_i$, $\tau_i$ and $\nu_i$ are the propagation gain, delay and Doppler-shift associated with $i$-th path. We assume that the delay-Doppler channel response is given by
\begin{equation}
h(\tau,\nu)=\sum_{i=1}^{P}h_i\delta(\tau-\tau_i)\delta(\nu-\nu_i)
\end{equation}
The delay-time channel in terms of the delay-Doppler response is given as
\begin{equation}
    g(\tau,t)=\int_\nu h(\tau,\nu){\rm e}^{j2\pi\nu(t-\tau)} d\nu    
\end{equation}
The received time-domain signal can then be written as
\begin{equation}
    r(t)=\int_\tau g(\tau,t)s(t-\tau) d\tau+w(t)    
\end{equation}
where $w(t)$ is the AWGN noise signal. The equivalent discrete-time channel is obtained by sampling the received time-domain waveform at $M\Delta f$ Hz, is given as
\begin{equation}
    {\bf r}[q]=\sum_{l \in \mathcal{L}} \thickbar{g}[l,q]{\bf s}[q-l]+{\bf w}[q]\label{TD_io}
\end{equation}
where $\tau=l/M\Delta f$, $t=q/M\Delta f$ and $\mathcal{L}$ denotes the set of discrete integer delay taps. The entries of $\thickbar{g}[l,q]$ is given as
\begin{equation}
    \thickbar{g}[l,q]=\sum_{i=1}^{P}h_i z^{\kappa_i(q-l_i)}\delta[l-l_i]\label{eq:TD_DD}
\end{equation}
where $z=e^{\frac{j2\pi}{NM}}$, $l_i=\frac{\tau_i}{M\Delta f}$ and $\kappa_i=\frac{\nu_i}{NT}$ are the {\em normalized delay} and {\em normalized Doppler-shift} associated with the $i$-th path. We assume that the {\em normalized delays} $l_i$ are integers.
\subsection{USC Receiver}
The transmitter operations are then reversed at the receiver. The delay-time received samples are obtained by folding the received signal ${\bf r}$ into a $M \times N$ matrix by filling it column-wise.
\begin{equation}
    \thicktilde{\bf Y}={\rm vec}^{-1}_{M,N}({\bf r}) \label{rx_step_1}
\end{equation}
The received information symbols can then be obtained as
\begin{equation}
    {\bf Y}=\thicktilde{\bf Y}\cdot{\bf U}_T^{\dag} \label{rx_step_2}
\end{equation}
The operations in (\ref{rx_step_1}) and (\ref{rx_step_2}) can be combined as 
\begin{equation}
    {\bf Y}[m,n]=\sum_{k=0}^{N-1}{\bf r}[m+kM]{\bf U}_{\rm T}^{\dag}[k,n] \label{DZT_form}
\end{equation}
Note that for ${\bf U}_{\rm T}={\bf F}_N^{\dag}$, the operation in (\ref{DZT_form}) is known in the literature as the discrete Zak transform (DZT) and is equivalent to the OTFS receiver for rectangular pulse shaping waveform, \cite{Journ_OTFS,Saif}.
Let the received 2-D symbol matrices ${\bf Y}$ and $\thicktilde{\bf Y}$ be vectorized as 
\begin{equation}
    {\bf y}={\rm vec}({\bf Y}^{\rm T}), \quad\thicktilde{\bf y}={\rm vec}(\thicktilde{\bf Y}^{\rm T})
\end{equation}
The received vector can then be written in terms of the transmitted vector as
\begin{equation}
    {\bf y}=({\bf I}_M \otimes {\bf U}_{\rm T}^{\dag})\cdot({\bf P}^{\rm T}\cdot{\bf r})=({\bf I}_M \otimes {\bf U}_{\rm T}^{\dag})\cdot\thicktilde{\bf y} \label{rx_perm}
\end{equation}
\subsection{USC input-output relations}
The input-output relation in (\ref{eq:TD_DD}) can be written in the matrix form as
\begin{equation}
    {\bf r}={\bf G}\cdot{\bf s}+{\bf w} \label{DT_io}
\end{equation}
where ${\bf G}$ is the time-domain channel matrix with a band width of $l_{\rm max}+1$ with entries: ${\bf G}[q,q-l]=\bar{g}[l,q]\text{ for }q \geq l$. 
Substituting (\ref{tx_perm}) and (\ref{rx_perm}) in (\ref{DT_io}), we get the input-output relation between the transmitted and received information symbols:
\begin{equation}
    {\bf y}={\bf H}\cdot{\bf x}+{\bf z}
\end{equation}
where ${\bf z}=({\bf I}_M\otimes{\bf U}_T^{\dag})\cdot({\bf P}^{\rm T}\cdot{\bf w})$ is the AWGN noise and the channel matrix 
\begin{align}
{\bf H}=({\bf I}_M\otimes{\bf U}_T^{\dag})\cdot({\bf P}^{\rm T}\cdot{\bf G}\cdot{\bf P})\cdot({\bf I}_M\otimes{\bf U}_T)
\end{align}

\section{Embedded pilot-aided channel estimation \label{sec:ch_est}}
In this section we present a time-domain channel estimation method for USC waveforms with ZP between blocks. Following the pilot placement in \cite{Ravi:ch_est,OTSM_journ}, a single pilot is embedded in the 2-D information symbol matrix ${\bf X}$ at location $(m_{\rm p},n_{\rm p})$. Guard symbols are placed around the single pilot to avoid interference between data and pilot.
\begin{align*}
    {\bf X}(m,n)= \left\{ \begin{array}{cc}
    x_{\rm p}& \text{if } m=m_{\rm p},n=n_{\rm p}\\
    0& \text{if } |m-m_{\rm p}| \leq l_{\rm max}\\
    \text{data symbols}& \text{otherwise.}
    \end{array}
    \right.  ~~~
\end{align*}
After transforming the information symbols to the delay-time domain, the matrix $\thicktilde{\bf X}$ containing pilot and data samples can be written as 
\begin{align*}
    \thicktilde{\bf X}[m,n]= \left\{ \begin{array}{cc}
    x_{\rm p}{\bf U}_{\rm T}[n_{\rm p},n]& \text{if } m=m_p\\
    0& \text{if } |m-m_{\rm p}| \leq l_{\rm max},\\
    \text{data samples}& \text{otherwise.}
    \end{array}
    \right.  ~~~
\end{align*}
The transmitted time domain vector can be written as the superposition of ${\bf s}_{\rm d} \in \mathbb{C}^{NM \times 1}$ containing the data samples and 
${\bf s}_{\rm p}\in \mathbb{C}^{NM \times 1}$ contains only the pilot samples:
\begin{equation}
    {\bf s}={\bf s}_{\rm d}+{\bf s}_{\rm p}
\end{equation}
where
\begin{align}
    {\bf s}_{\rm p}[m+nM]= \left\{ \begin{array}{cc}
    x_{\rm p}{\bf U}_{\rm T}[n_{\rm p},n]& \text{if } m=m_{\rm p}\\
    0&  \text{otherwise}
    \end{array}
    \right.  ~~~
\end{align}
The interference between the data and pilot samples are avoided due to the guard samples between the data and pilot. This allows the receiver to process the pilot samples for channel estimation separately from data for detection.

From (\ref{TD_io}), the received pilot samples are related to the transmitted pilot samples as
\begin{align}
    {\bf r}_{\rm p}[q+l]=&\bar{g}[l,q+l]{\bf s}_{\rm p}[q]+{\bf w}[q+l]
\end{align}
where ${\bf s}_{\rm p}[q]=0$ for $q \neq m_{\rm p}+nM$. The time domain channel coefficients can then be estimated using the pilot samples at locations $q=(m_{\rm p}+nM)$ for $n=0,\ldots,N-1$ and $l=0,\ldots,l_{\rm max}$ as
\begin{align}
    \hat{\bar{g}}[l,m_{\rm p}+nM&+l]={\bf r}_{\rm p}[m_{\rm p}+nM+l]/{\bf s}_{\rm p}[m_{\rm p}+nM]
    \nonumber \\
    &={\bf r}_{\rm p}[m_{\rm p}+nM+l]/(x_{\rm p}{\bf U}_{\rm T}[n_p,n])
    \label{ch_est:est}
\end{align}
The time-domain channel for the entire frame can then be obtained by interpolating the estimated time domain channel coefficients at locations $(m_{\rm p}+nM+l)$ in (\ref{ch_est:est}). The estimated channel coefficients $\hat{\bar{g}}[l,q]$ can be imagined as the delay-time channel sub-sampled by a factor $M$. Since the time-domain channel for each delay tap can be modelled as the sum of sinusoids (see (\ref{eq:TD_DD})) corresponding to Doppler-shift of the paths in the delay bin, spline or linear interpolation can be used to reconstruct the delay-time channel coefficients for the entire frame. Successful reconstruction is possible as long as the sampling rate of the delay-time channel ($\Delta f$) is at least twice the maximum frequency component (Doppler-shift) of the delay-time channel, i,e., $\nu_{\rm max}<\Delta f/2$, which is a reasonable assumption for the typical mobile wireless channels.   

\section{Low complexity detection}
We consider the case when ZPs are inserted between the time-domain blocks since the ZP can be used to embed pilots as shown in the previous section. In this case, the input output relation in (\ref{DT_io}) can be split into $N$ independent blocks as
\begin{equation}
    {\bf r}_n={\bf G}_n\cdot{\bf s}_n+{\bf w}_n, \quad n=0,\ldots,N-1 \label{time_domain_io}
\end{equation}
where ${\bf G}_n \in \mathbb{C}^{M\times M}$ and ${\bf w}_n\in \mathbb{C}^{M\times 1}$ are the time-domain channel matrix and the zero-mean noise vector with covariance vector ${\sigma_{\rm w}^2}{\bf I}_{M}$, respectively, in the $n$-th time slot.
\subsection{Frequency-domain single tap equalizer\label{sec:single_tap_eq}}
For fair comparison with the traditional OFDM modulation scheme, we use the single tap frequency domain equalizer for USC waveforms. The received time-frequency samples can be obtained by the $M$-point FFT operation on the received time domain blocks
\begin{align}\thickbar{\bf r}_n={\bf F}_M\cdot{\bf r}_n.\label{ytf}\end{align}
as in a standard OFDM receiver.
We can then equalize each block in parallel as
\begin{equation}
    \thickbar{\bf s}_n(m)=\frac{\thickbar{\bf h}_n^{\ast}[m]\cdot\thickbar{\bf r}_n[m]}{\lvert \thickbar{\bf h}_n[m]\rvert^{2}+\sigma_w^2}, \quad m=0,\ldots,M-1
    \label{sn}
\end{equation}
for $n=0,\ldots,N-1$ and the frequency domain channel coefficients for each time-domain block are given by
\begin{equation}
    \thickbar{\bf h}_n={\rm diag}[{\bf F}_M\cdot{\bf G}_n\cdot {\bf F}_M^{\dag}]
    \label{diag}
\end{equation}
where ${\rm diag}[{\bf A}]$ denotes the column vector containing the diagonal elements of the square matrix ${\bf A}$. 
The information symbol estimates in the delay-sequency domain can then be obtained by the $M$-point IFFT operation on the time-frequency domain estimates $\thickbar{\bf s}_n$ followed by the $N$-point WHT as
\begin{equation}
    \hat{\bf X}={\bf F}_M^{\dag}\cdot[\thickbar{\bf s}_0, \thickbar{\bf s}_1, \ldots, \thickbar{\bf s}_{N-1}]\cdot{\bf U}_{\rm T}^{\dag}\label{est_singletap}
\end{equation}
\begin{figure}
    \centering
    {\includegraphics[trim=10 0 0 10,clip,height=1.9in,width=3.2in]{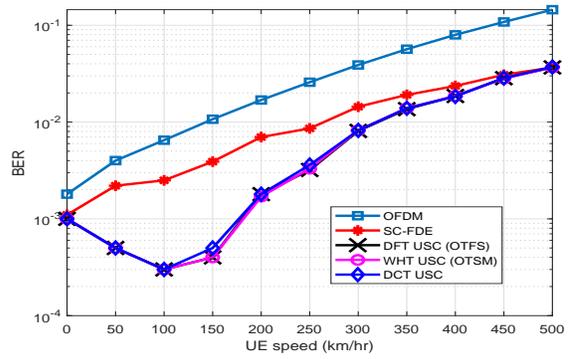}
     \vspace{-3mm}\caption{BER performance of single tap equalizer with QPSK for WHT (OTSM), DFT (OTFS) and DCT compared with OFDM for different speeds at SNR = 20dB}
    \label{single_tap}}
    \end{figure}
    Fig.~\ref{single_tap} shows the QPSK BER performance of USC schemes compared with SC and OFDM modulation at a SNR of 20 dB for different speeds. We consider three different USC waveforms with DFT (OTFS). WHT (OTSM) and DCT. It can be observed that all the USC waveforms offer better performance as compared to OFDM and SC. However, as expected, the time-frequency single tap equalizer performance degrades in the presence of significant Doppler due to significant inter-carrier interference.

\subsection{Time-domain linear minimum mean-squared error equalizer}
The frequency-domain equalizer performance, similar to OFDM, degrades in the presence of high Doppler spread channels (see Fig. \ref{single_tap}). In this section we present a block-wise time-domain MMSE equalizer that can provide better performance in the case of time-varying channels. The MMSE equalizer will act as a baseline to compare the performance of the iterative detection methods we present later.

For the input-output relation in (\ref{time_domain_io}), the MMSE estimate of the time-domain samples,
\begin{align}
    \hat{{\bf s}}_n=({\bf G}_n^{\dag}\cdot{\bf G}_n+\sigma_w^{2}{\bf I}_M)^{-1}\cdot{\bf G}^{\dag}\cdot{\bf r}_n
\end{align}
The information symbols can then be estimated from the delay-time samples as 
\begin{equation}
    \hat{\bf X}=[\hat{\bf s}_0, \hat{\bf s}_1, \ldots, \hat{\bf s}_{N-1}]\cdot{\bf U}_{\rm T}^{\dag} \label{est_MMSE}
\end{equation}

\subsection{Time-domain matched-filtered Gauss Seidel equalizer}
 The time-domain MMSE equalizer offers good performance, but at the cost of high complexity. To reduce the complexity, we present an iterative detector based on the well known Gauss Seidel (GS) method, \cite{LSBook}. However, different from the traditional method, the GS iteration in this case is done on the matched filtered channel matrix blocks ${\bf R}_n={\bf G}^{\dag}_n\cdot{\bf G}_n$.
 The matrix input-output relation in (\ref{time_domain_io}) after the matched filtering operation can be written as 
 \begin{equation}
    {\bf z}_n={\bf R}_n\cdot{\bf s}_n+\thickbar{\bf w}_n
    \label{MRC_matrix}
\end{equation} 
where ${\bf R}_n={\bf G}^{\dag}_n\cdot{\bf G}_n$, ${\bf z}_n={\bf G}^{\dag}_n\cdot{\bf r}_n$ and $\thickbar{\bf w}_n={\bf G}^{\dag}_n\cdot{\bf w}_n$.
The GS method is used to iteratively find the least squares solution
\begin{equation}
   \hat{\bf s}_n=\min_{{\bf s}_n} ||{\bf z}_n-{\bf R}_n{\bf s}_n||^2
\end{equation}
of the $M$-dimensional linear system of equations in (\ref{MRC_matrix}).

 Let ${\bf D}_n$ and ${\bf L}_n$ be the matrix containing the diagonal elements and the strictly lower triangular elements of the matched filter matrix ${\bf R}_n$. From \cite{LSBook}, the GS iterative method for finding the estimate of ${\bf s}_n$ in each iteration is given as 
\begin{align}
    &\hat{\bf s}_n^{(i)}=-{\bf T}_n\cdot\hat{\bf s}_n^{(i-1)}+{\bf b}_n  \\
&{\bf T}_n=({\bf D}_n+{\bf L}_n)^{-1}\cdot{\bf L}_n^{\dag},\quad{\bf b}_n=({\bf D}_n+{\bf L}_n)^{-1}\cdot{\bf z}_n \label{Tn}
\end{align}
where ${\bf T}_n \in \mathbb{C}^{M \times M}$ is the GS iteration matrix. The vector $\hat{\bf s}_n^{(i)} \in \mathbb{C}^{M \times 1}$ represents the estimate of the transmitted time-domain samples of the $n$-th block in the $i$-th iteration. The information symbols in the $i$-th iteration is then given as 
\begin{equation}
    \hat{\bf X}^{(i)}=\mathcal{D}\left({\bf C}^{(i)}\right),\text{ where }{\bf C}^{(i)}=[\hat{\bf s}_0^{(i)}, \hat{\bf s}_1^{(i)}, \ldots, \hat{\bf s}_{N-1}^{(i)}]\cdot{\bf U}_{\rm T}^{\dag} \label{demod}
\end{equation}
where $\mathcal{D}(.)$ denotes the decision function replacing all the elements of  the input with the nearest QAM symbol. The hard decision estimates are transformed back to the time domain to update the time domain estimate to be used in the next iteration. 
\vspace{-2mm}
\begin{equation}
    \hat{\bf s}^{(i)}\leftarrow(1-\delta)\hat{\bf s}^{(i)}+\delta{\rm vec}\left({\bf X}^{(i)}\cdot{\bf U}_{\rm T}\right) \label{mod}
\end{equation}
where $\delta$ is the relaxation parameter to improve the detector convergence for higher modulation schemes like 64-QAM, \cite{LSBook,Journ_OTFS}. As initial estimate to the iterative detection, we can chose either ${\bf X}^{(0)}={\bf 0}$ or the single-tap solution in (\ref{est_singletap}) yielding faster convergence.  Note that the complexity of this algorithm can be significantly reduced by taking advantage of the sparsity of the matrices ${\bf G}_n$ and ${\bf T}_n$. It is shown in \cite{Journ_OTFS}, that a delay-time version of this algorithm for OTFS has complexity $O(NML)$, where $L$ is the number of distinct channel delays. The same implementation can be straightforward extended to other USC waveforms.

\section{Results and Discussion}
For all simulations, we consider QPSK and a frame size of $N=M=64$. The sub-carrier spacing of 15 kHz is used and the carrier frequency is set to 4 GHz.
 The maximum delay spread (in terms of integer taps) is taken to be 4 ($l_{\max}=3$) which is approximately 4 ${\mu}s$, i.e., $L=4$. The channel delay model is generated according to the standard EVA model
 with the Doppler shift for the $i$-th path $\nu_i=\nu_{\max}\cos(\theta_i)$ with $\theta_i$ generated from the uniform distribution $U(-\pi,\pi)$, where $\nu_{\max}$ is the maximum Doppler shift \cite{EVA}. The channel is estimated as described in Section \ref{sec:ch_est}. For every SNR point in the BER plots, $10^5$ frames are simulated.
 
\begin{figure}
    \centering
    {\includegraphics[trim=10 0 0 10,clip,height=1.9in,width=3.2in]{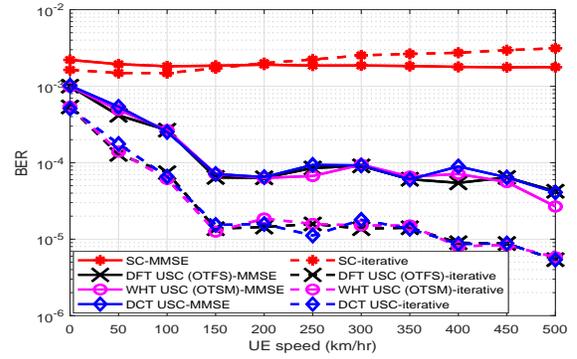}
     \vspace{-3mm}\caption{ BER performance of MMSE and iterative equalizer for different speeds at SNR=20 dB}
    \label{fig:comp_1}}
    \end{figure}

\begin{figure}
    \centering
    {\includegraphics[trim=10 0 0 10,clip,height=1.9in,width=3.2in]{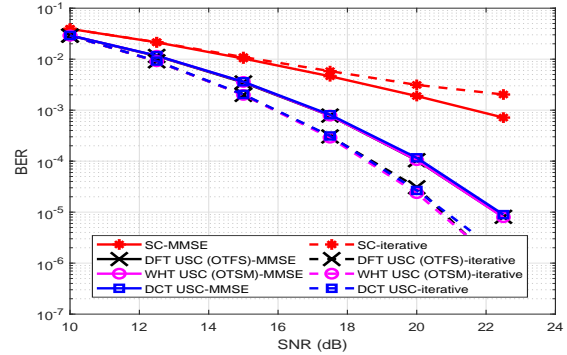}
     \vspace{-3mm}\caption{BER performance with iterative equalizer for OTSM, OTFS, DCT-OTFS compared with SC scheme at 500 km/hr}
    \label{fig:comp_2}}
    \end{figure}

 Fig. \ref{fig:comp_1} shows the BER performance for USC waveforms with DFT (OTFS), WHT (OTSM) and DCT unitary matrix compared with the SC scheme for different UE speeds at 20 dB SNR. It can be observed that as the UE speed increases, the BER performance of the USC schemes improves with increasing Doppler spread where as the SC waveform does not gain from the available Doppler diversity. Further the low-complexity iterative equalizer offers better performance than the high-complexity MMSE equalizer. Fig. \ref{fig:comp_2} shows the BER performance of the USC schemes compared with SC using MMSE and iterative equalizers. It can be observed that the USC scheme for any choice of the unitary matrix ${\bf U}_{\rm T}$ offers around $\approx 6$ dB gain at $10^{-3}$ BER over SC at a UE speed of 500 km/hr.


\subsection{Real-time indoor experiment using SDR} 
The hardware platform is based on National Instruments universal radio software peripheral (USRP) software defined radio reconfigurable device (NI-USRP-2954R) designed by Ettus Research \cite{USRP}. 
A Tx or Rx terminal is implemented with an USRP-2954R connected to a host PC running the National Instruments LabView using PCIe Express x4. The software is based on LabView 2020. We follow the OTFS SDR modem experimental setup described in \cite{SDR} with the USC experiment parameters listed in Table \ref{tab1}.

\begin{figure}
    \centering
    {\includegraphics[trim=10 0 0 10,clip,height=2 in,width=3.2in]{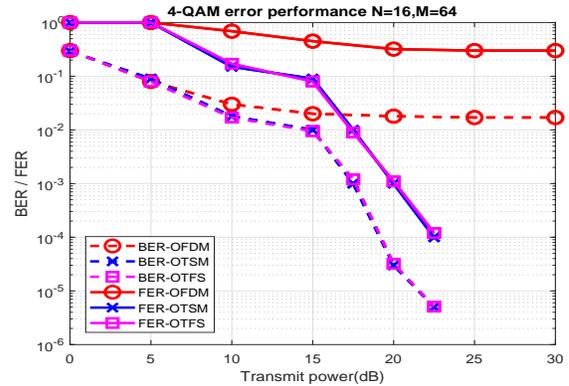}
     \vspace{-3mm}\caption{4-QAM BER and FER performance of OTSM, OTFS and OFDM using single tap equalizer in real indoor channel}
    \label{fig:BER_SDR}}
    \end{figure}

\begin{table}
\renewcommand{\arraystretch}{1.3}
\centering
\caption{Experiment Parameters}
  \label{tab1}
  \begin{tabular}[c]{ | c | M{5cm} | c | }
    \hline
    Symbol & Parameter & Value \\\hline
    $f_c$ & Carrier frequency & 4 GHz\\\hline
    $M \times N$ & Number of subcarriers $\times$ time-slots & $64 \times 64$ \\\hline
    $T$ & Symbol Time & 32 $\mu$s\\\hline
    $\Delta f$ & Subcarrier spacing & 31.25 KHz\\\hline
    $d$ & Tx-Rx Distance & 5 meters  \\\hline
  \end{tabular}
  \vspace{-4mm}
\end{table}
\begin{figure}
    \centering
    {\includegraphics[trim=10 10 10 10,clip,height=1.9in,width=3in]{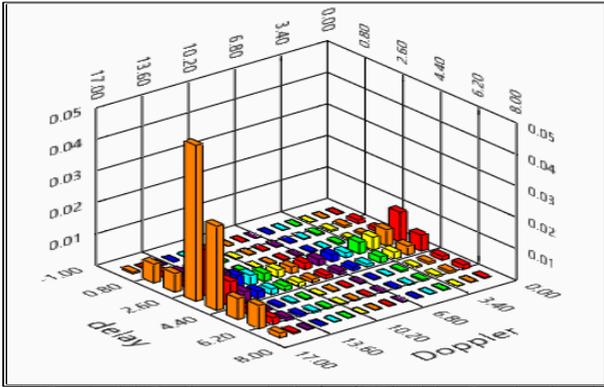}
     \vspace{-3mm}\caption{Received 2-D pilot magnitude for OTFS (delay-Doppler domain) in the indoor wireless channel}
    \label{fig:SDR_DD}}
    \end{figure}
    \begin{figure}
    \centering
    {\includegraphics[trim=10 10 10 10,clip,height=1.9in,width=3in]{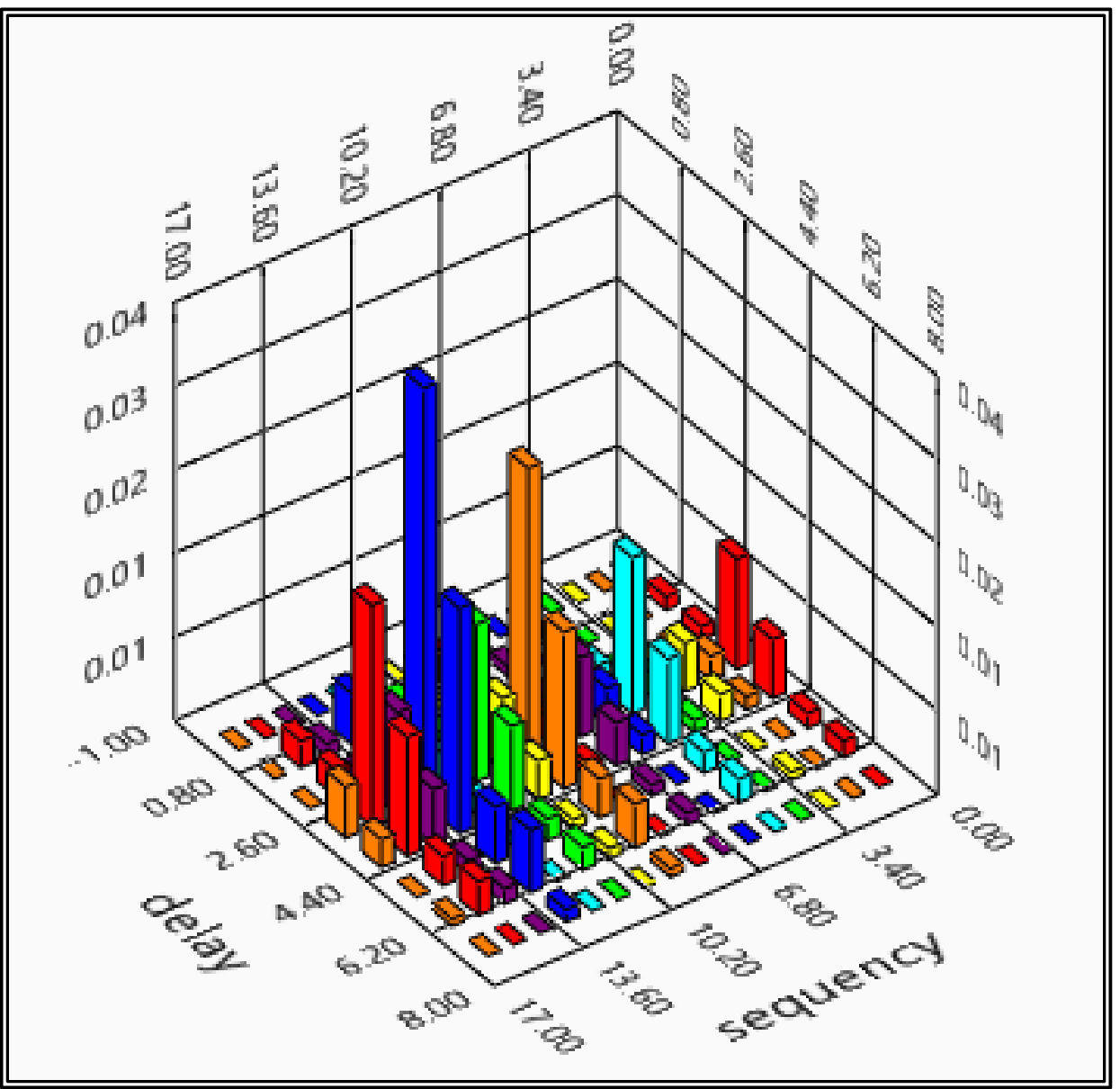}
     \vspace{-3mm}\caption{Received 2-D pilot magnitude for OTSM (delay-sequency domain) in the indoor wireless channel}
    \label{fig:SDR_DS}}
    \end{figure}
Fig. \ref{fig:BER_SDR} presents the BER and FER performance of USC with DFT (OTFS) and WHT (OTSM) unitary matrices compared with OFDM in a real-time indoor channel using the NI USRP-2954 SDR platform. The OFDM frame of bandwidth $B=2$ MHz is generated according to the IEEE 802.11ac standard for WiFi using 48 out of 64 sub-carriers for data transmission and the rest for pilot and null symbols. The pilot overhead in USC waveforms is generated with a ZP of $L_{\rm G}=16$ samples to match the spectral efficiency of above OFDM. The information bits are encoded using a convolutional code of rate $R=1/2$. At the Rx the single-tap equalizer in Section \ref{sec:single_tap_eq} is used for low-complexity detection and, the channel estimation method is given in Section \ref{sec:ch_est}. 

An instance of the received 2-D pilot magnitude in the indoor channel measured using the SDR is shown in Figs. \ref{fig:SDR_DD} and \ref{fig:SDR_DS} for OTFS and OTSM for a pilot transmitted at $(m_{\rm p},n_{\rm p})=(3,0)$. It can be observed that the delay-Doppler domain channel is more localized than the delay-sequency channel. However, the pilot energy is still leaked to all the Doppler bins due to fractional Doppler. Therefore, the entire Doppler axis needs to be reserved for the guard symbols to avoid interference with data. From  Fig. \ref{fig:BER_SDR}, both both OTFS and OTSM offer better performance than OFDM in the indoor channel.

One reason for a high error floor in OFDM is due to the lack of CFO compensation at the receiver. OTSM, similar to OTFS, is more robust to CFO as compared to OFDM. Due to the significantly lower complexity of the WHT compared to the DFT and other unitary matrices, Moreover, if the information symbols are integers, the time domain samples after applying the WHT resulting in the least quantization error at the transmitter for the same DAC resolution. Due to above reasons, OTSM can be considered a more energy-efficient waveform than OFDM and OTFS for next-generation wireless channels.
\section{Conclusion}
In this paper, we defined a family of waveforms based on precoding the time dimension of the delay-time domain using unitary matrices. We presented channel estimation and detection methods for USC waveforms. Any choice of unitary matrix was shown to offer the same performance as the recently proposed OTFS modulation, but differs in cost of implementation depending on the unitary transform complexity. Finally we verify the performance of USC waveform in a real-time indoor channel using an SDR platform.

\end{document}